\documentstyle[epsfig]{aa}

\def\ltsima{$\; \buildrel < \over \sim \;$}
\def\simlt{\lower.5ex\hbox{\ltsima}} 
\def\gtsima{$\; \buildrel > \over \sim \;$}
\def\simgt{\lower.5ex\hbox{\gtsima}} 

\makeatletter
\def\@cite#1#2{(#1\if@tempswa , #2\fi)}
\def\@citex[#1]#2{\if@filesw\immediate\write\@auxout{\string\citation{#2}}\fi
  \def\@citea{}\@cite{\@for\@citeb:=#2\do
    {\@citea\def\@citea{;\penalty\@m\ }\@ifundefined
       {b@\@citeb}{{\bf ?}\@warning
       {Citation `\@citeb' on page \thepage \space undefined}}%
\hbox{\csname b@\@citeb\endcsname}}}{#1}}

\begin{document}


\title{BeppoSAX confirmation of beamed afterglow emission from GRB990510}

\author{Elena Pian\inst{1,2}
\and Paolo Soffitta\inst{3}
\and Alessandro Alessi\inst{4}
\and Lorenzo Amati\inst{2}
\and Enrico Costa\inst{3}
\and Filippo Frontera\inst{5,2}
\and Andrew Fruchter\inst{6}
\and Nicola Masetti\inst{2}
\and Eliana Palazzi\inst{2}
\and Alin Panaitescu\inst{7}
\and Pawan Kumar\inst{8}} 

\offprints{Elena Pian, pian@tesre.bo.cnr.it}

\institute{Osservatorio Astronomico di Trieste, Via G.B.
Tiepolo 11, I-34131 Trieste, Italy
\and
Istituto Tecnologie e Studio Radiazioni Extraterrestri, CNR,
Via Gobetti 101, I-40129 Bologna, Italy
\and
Istituto di Astrofisica Spaziale, CNR, via Fosso del
Cavaliere, Area di Ricerca Tor Vergata, I-00133 Rome, Italy
\and
Dipartimento di Astronomia, Universit\`a di Bologna, Via Ranzani 1,
I-40127 Bologna, Italy
\and
Universit\`a di Ferrara Dipartimento di Fisica, Via
Paradiso 11, I-44100 Ferrara, Italy
\and
Space Telescope Science Institute, 3700 San Martin
Drive, Baltimore, MD 21218, USA
\and
Department of Astrophysical Sciences, Princeton University, NJ 08544, USA
\and
Institute for Advanced Study, Olden Lane, Princeton, NJ 08540, USA
}

\date{Received;  Accepted }


\markboth{E. Pian et al.: BeppoSAX confirmation of beamed afterglow
emission from GRB990510}{}


\abstract{
We compare the prompt X-ray (2-10 keV) emission of GRB990510 measured
by the BeppoSAX Wide Field Cameras (WFC) during the burst to the X-ray
afterglow detected by the BeppoSAX Narrow Field Instruments.  A single
power-law model for the afterglow, $f(t) \propto t^{-1.42}$, is ruled
out.  Provided the initial time of the afterglow is assumed to coincide
with the last short pulse in the X-ray prompt event
(i.e., 72
seconds after the GRB trigger time), the X-ray emission from $\sim$80
to $10^5$ seconds after the GRB trigger is well described by an external shock
expanding in a decelerating jet, in which synchrotron radiation takes
place. This model, represented by a double power-law of indices
$\alpha_1 \simeq 1$ and $\alpha_2 \simeq 2$ before and after a jet
collimation break time of $\sim$0.5 days after GRB, respectively, is
consistent with the second and third upper limits measured by the WFC,
but not with the first. This may be related to inhomogeneities in the
circumburst medium.  Our finding indicates that the temporal behavior
of the GRB990510 X-ray afterglow is similar to that at optical
wavelengths, and thus strengthens the interpretation of the
multiwavelength afterglow as synchrotron emission in a jet with
decreasing Lorentz factor.  GRB990510 is thus the only burst in which
evidence of a spreading jet has been found in X-rays.
}


\maketitle


\begin{figure*}
\begin{center}
\psfig{figure=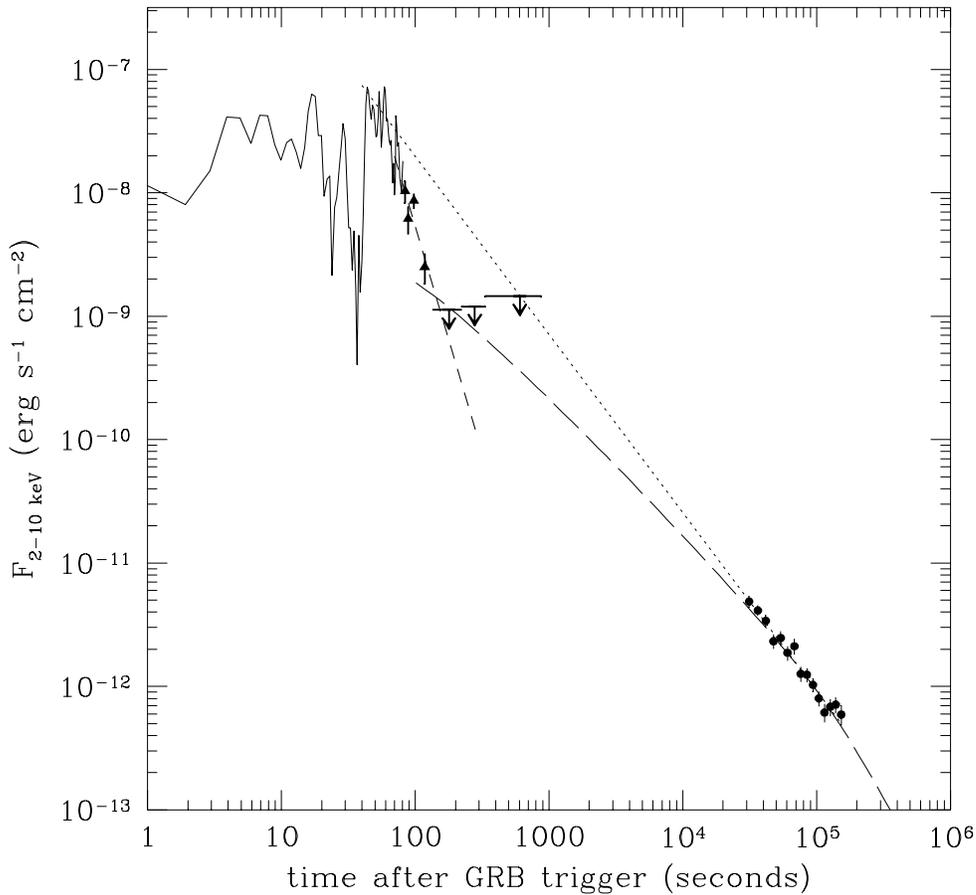,width=14cm}
\end{center}

\caption[]{BeppoSAX light curves of GRB990510 in the 2-10 keV range: WFC
temporal profile (solid curve, filled triangles and 3-$\sigma$ upper
limits) and afterglow NFI measurements (filled circles). Also shown are
the single power-law $f(t)  \propto t^{-1.42}$ (dot) fitted to the
afterglow by KAK00, the power-law $f(t) \propto
t^{-3.7}$ (short dash) which best-fits the last four points of the WFC
profile, and the double power-law of indices $\alpha_1 \simeq 1$ and
$\alpha_2 \simeq 2$ (long dash) derived from PK00 afterglow
model ($z = 1.62$. The fit has $\chi^2 = 41$ for 69 degrees of freedom). 
The six model parameters are isotropic equivalent energy $E_0 = 5
\times 10^{53}$ erg, initial jet half-opening angle $\theta_0 = 2.7$
degrees, density of homogeneous external medium $n = 0.13$ cm$^{-3}$,
fraction of shock energy going into electrons $\epsilon_e = 4.6
\times 10^{-2}$, magnetic energy density $\epsilon_B = 9
\times 10^{-4}$, exponent of injected power-law electron distribution $p =
2$.  The amount of jet energy lost radiatively until 2 days after GRB
is 45\%.  Note that these parameters, slightly revised with respect to
those reported in PK00 for this burst, are affected by
rather large uncertainties, and must therefore only be considered as
representative. The extrapolation of the single power-law $t^{-1.42}$
backward to the time of the prompt event is not consistent
with the WFC latest points and upper limits, while the double power-law
matches well the last WFC point and upper limits}

\end{figure*}

\begin{figure*}
\begin{center}
\psfig{figure=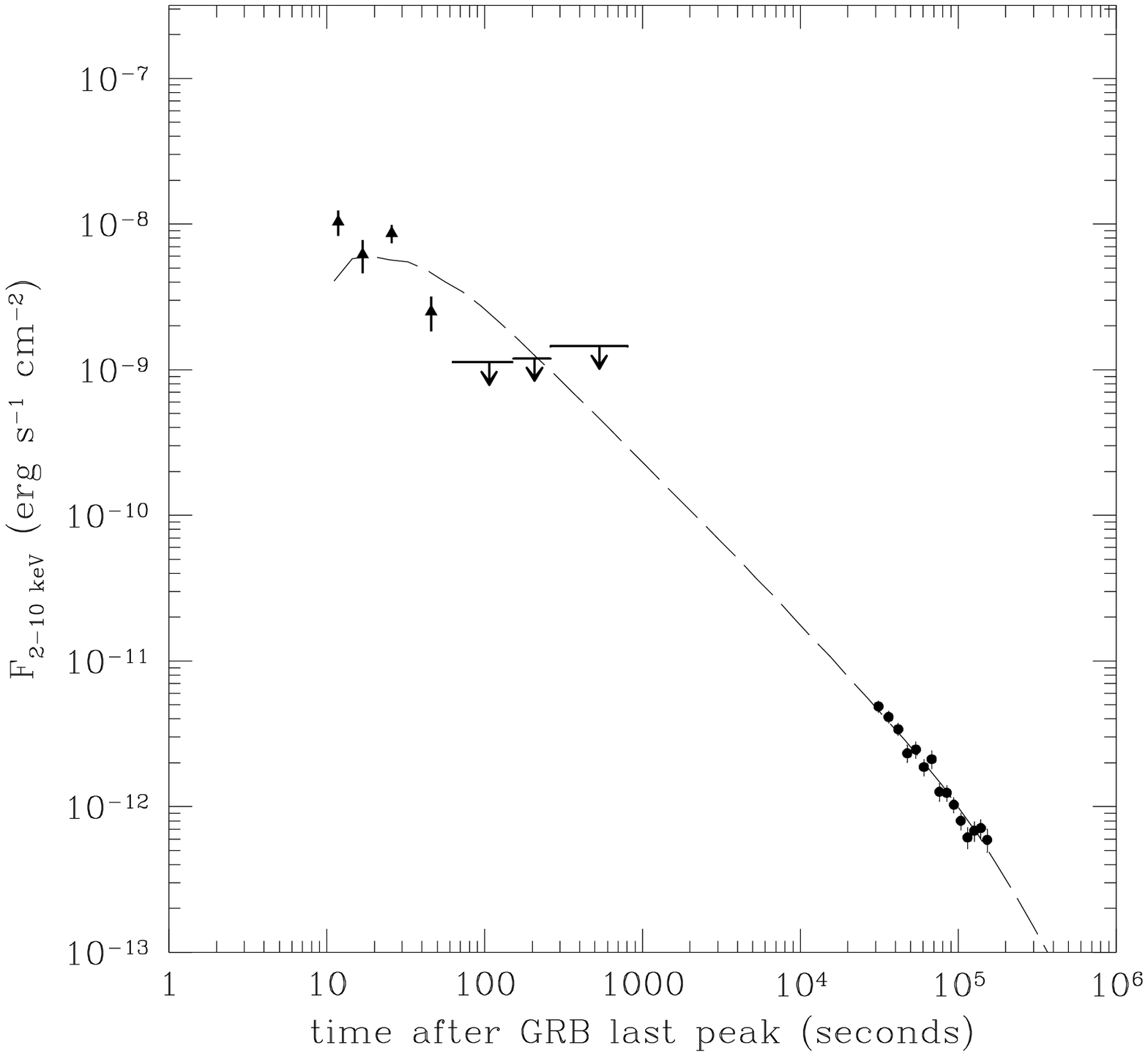,width=14cm}
\end{center}

\caption[]{BeppoSAX WFC (last four measurements and three upper limits) and
NFI light curves of GRB990510 in the 2-10 keV range, as in Fig. 1, but with
assumed initial time equal to the time of the last GRB peak (i.e., 72 seconds
after trigger, see Fig. 1). Overimposed is the curve obtained by fitting the
PK00 model to the multiwavelength afterglow data, plus the 4 last WFC points. 
The downward bending of the model curve toward the earliest epochs is due to
the fact that the observer frame deceleration timescale of the blast implied
by the model parameters is $\sim$6 seconds, so that X-ray emission starts
rising only after this time}

\end{figure*}

\section{Introduction}

GRB990510 was detected by the CGRO BATSE instrument and by the Gamma-Ray
Burst Monitor (GRBM) onboard BeppoSAX on 1999 May 10.3674 UT.  Following
the localization by the BeppoSAX Wide Field Cameras (WFC, 2-26 keV) Unit 2,
the afterglow search promptly started and allowed the detection of a bright
and variable source at X-ray, optical and radio wavelengths (Kuulkers et
al.  2000, hereafter KAK00; Harrison et al. 1999; Stanek et al. 1999; Israel
et al. 1999). 
Polarimetry of the optical transient yielded the first measurement of an
afterglow polarization percentage (1.7\%, Wijers et al.  1999; Covino et
al. 1999).  Optical spectroscopy detected many absorption lines and led to
the estimate of a lower limit on the redshift, $z > 1.62$ (Vreeswijk et al. 
2001). Late epoch photometry from ground and with HST put limits on the
magnitude of the host galaxy (Beuermann et al. 1999;  Fruchter et al. 
1999)  and eventually provided evidence for a faint, small nebulosity
(Fruchter et al. 2000;  Bloom 2000).  The optical and radio temporal
behavior of the afterglow suggests that this has developed within a jet
progressively spreading and laterally expanding (Harrison et al. 1999;
Stanek et al. 1999).  Panaitescu and Kumar (2000, hereafter PK00)  have
modeled the
afterglow multiwavelength light curves with synchrotron radiation in a
relativistic jet expanding in a medium of uniform density. 

The signature of a decreasingly collimated jet is the simultaneous
steepening of afterglow light curves at all frequencies (Sari et al. 
1999;  Rhoads 1999; see however Wei \& Lu 2000).  This is clearly seen in
the optical data of GRB990510, which can be described by two temporal
power-laws, $f(t) \propto t^{-\alpha}$, of indices $\alpha \sim 0.8$ and
$\alpha \simgt 2$ before and after a break occurred at approximately 1 day
after the GRB, respectively.  Instead, the X-ray afterglow data are not
critical in discerning between a spherical or jet hydrodynamical expansion
(KAK00), due to limited signal-to-noise ratio of the
BeppoSAX Narrow Field Instruments (NFI) data, to the relatively long
interval elapsed between GRB detection and NFI pointing (8 hr), and to the
limited duration of the observation (36 hr).  The data are consistent both
with a single power-law $f(t) \propto t^{-1.42}$ and with a double
power-law. Based on the standard afterglow synchrotron model, KAK00 
argue that if decreasing beaming is taking place, a break should occur
at $\sim$1 day after the GRB, as in the optical, and the indices of the
temporal power-laws should be $\sim$1.1 before the break, i.e. steeper
than in the optical by $\Delta \alpha \simeq 0.25$, and $\sim$2.1 after
the break, i.e., equal to the optical (see Sari et al. 1998; Sari et al.
1999).  By applying these parameters to the X-ray light curve, KAK00 
find a good agreement. 

Since the two descriptions of the X-ray afterglow light curve (single
or double power-law) would imply a very different flux level at the
early epochs (from few to hundreds of seconds after the GRB), a crucial
test to select the correct model is provided by the comparison of the
X-ray afterglow NFI data and the X-ray prompt event WFC data. Moreover,
simultaneous multiwavelength spectra of the afterglow at many epochs
would allow us to follow its temporal evolution and specifically to map
the behavior of the synchrotron cooling frequency, which is critical in
distinguishing an isotropic from a beamed expansion.  Therefore, we
have retrieved from the BeppoSAX archive and analyzed the WFC data of
the prompt event, and we have collected from the literature
simultaneous optical and X-ray data of the afterglow of GRB990510.  Our
WFC data analysis is described in Sect. 2, and the results are compared to
the X-ray NFI afterglow data in Sect. 3. The multiwavelength data analysis
and results are reported in Sect. 4.  The X-ray and multiwavalength
findings are discussed in Sect. 5, and a summary is given in Sect. 6.

\section{Wide Field Camera data analysis and results}

The WFC data of the prompt GRB990510 event have been retrieved from the
BeppoSAX archive (see also Briggs 2000) and analyzed by means of a standard
package which includes the ``iterative removal of source"  procedures (IROS,
V. 105.108, e.g.  Jager et al. 1997). These implicitly subtract the background
and the other sources in the field of view.  The burst direction was offset by
$15^{\circ}$ with respect to the WFC axis, which implies an effective
collecting area of 32 cm$^2$. We have analyzed the spectrum measured by the
WFC during the burst using the XSPEC package (V. 10.0, Arnaud 1996), and
obtained a good fit with a power-law $f(\nu)  \propto \nu^{-\beta}$ of index
$\beta = 0.23 \pm 0.08$ (absorbed by Galactic neutral hydrogen, $N_{HI} = 0.94
\times 10^{21}$ cm$^{-2}$, see also KAK00).  We used this spectral shape to
convert the WFC count rates to fluxes. The burst fluence in the 2-10 keV
energy range is $(3.3 \pm 0.1) \times 10^{-6}$ erg cm$^{-2}$.

Fig.~1 shows the temporal profile of the prompt event and of the X-ray
afterglow measured by the NFI during the interval from 8 to 44 hours after the
burst in the energy range 2-10 keV.  Significant WFC signal is detected up to
$\sim$130 s after the GRB trigger. The four last points of the WFC
profile
(encompassing the interval from 80 to 130 s after GRB and shown as filled
triangles in Fig. 1) indicate a fast decay, best-fitted with a temporal
power-law of index $\alpha = 3.7 \pm 1.5$.  The spectrum measured by the WFC
during this time interval is well fitted with a power-law of slope $\beta =
0.96 \pm 0.26$.  This is consistent with the upper limits provided by the
BeppoSAX GRBM (40-700 keV)  in the same time interval. In Fig.~1 are also
reported the 3-$\sigma$ upper limits on the X-ray flux, determined by the WFC
after the GRB subsided under background.

\section{Comparison of the prompt and afterglow X-ray emission}

The tail of the X-ray prompt emission, as measured by the WFC, and the
following upper limits are inconsistent with an extrapolation of the single
power-law fitted by KAK00 to the afterglow back to the time of the burst, as
shown in Fig.~1.  Therefore the X-ray afterglow can not follow a single
power-law, and may instead have a shallower decay at earlier times and then
continuously steepen, similar to the optical behavior. To test this hypothesis
and to predict the X-ray afterglow flux at the early epochs after the GRB
($\sim$100 seconds), we have adopted the model of PK00, which accounts well
for the multiwavelength afterglow emission of GRB990510 (see their Fig. 3).

A fit of the model to the NFI X-ray, optical V and I band, and 8.5 GHz radio
data of the afterglow (i.e., past $10^4$ seconds after trigger) yields a
$\chi^2 = 41$ for a total of 69 degrees of
freedom.  The fit parameters are reported in the caption to Fig. 1.  The WFC
X-ray measurements and upper limits have not been included in the fit, because
we have made the hypothesis that the early X-ray emission is
due to other mechanisms, so that the basic modeling assumption, that a
self-similar outflow has set in, may not be satisfied.

The model curve asymptotically tends, at early and late epochs, to power-laws
of indices $\alpha_1 \sim 1$ and $\alpha_2 \sim 2$, respectively, with an
achromatic collimation break time of $\sim$0.5 days, compatible with that
empirically determined by Stanek et al. (1999), Harrison et al. (1999), Israel
et al. (1999).  The temporal indices are consistent with those suggested by
KAK00 for the X-ray afterglow in a beamed scenario, and in fact the curve
reproduces well the trend of the X-ray afterglow data. A comparison to the
prompt X-ray data shows that the afterglow model curve lies below the four
latest WFC measurements, but in excellent agreement with the last of them and
with the three WFC upper limits (Fig. 1). 

However, while this match in flux between the observed last point of the X-ray
prompt event and the predicted afterglow at that epoch is very good (and
supported by the WFC upper limits), the change in temporal slope between the
tail of the prompt event and the early afterglow is large (from $\alpha \sim
3.7$ to $\alpha \sim 1$).  Therefore, we have considered the possibility that
this abrupt change is only determined by the choice of the initial time, and
we have assumed that the X-ray afterglow starts at the time of the last
peak of the WFC profile, occurring at 72 seconds after the GRB trigger
time (see
Fig.~1). This assumption is more physical, because one expects that different
pulses in a GRB light curve are produced by different successive shocks,
and
that the afterglow sets in after the last one.  Under this assumption, the
power-law index of the last 4 WFC points temporal decay is $\alpha = 0.9 \pm
0.2$ (with a reduced $\chi^2_\nu = 4.5$, due to the high scatter of the
points).  In Fig.~2 we report the last 4 WFC points, the 3 upper limits and
NFI data at epochs computed with respect to the GRB last peak.  We have
re-fitted the PK00 model to the multiwavelength data, by adding this time the
4 last WFC fluxes (but not the 3 upper limits), which are now assumed to
be part of the afterglow, and
therefore produced by an external shock.  The fit has $\chi^2 = 50$ for 73
degrees of freedom, and the model curve is reported in Fig. 2, superimposed
onto the data. The new fit parameters are very close to the previous ones,
obtained by assuming a same initial time for GRB and afterglow, and by
excluding the 4 latest WFC measurements.  The model curve accounts well for
all WFC and NFI data, except the first of the 3 WFC upper limits, which lies
below it by a factor $\sim$2. 

\section{The broad-band spectrum}

In a spreading jet scenario the synchrotron cooling frequency, $\nu_C$, is
predicted to decrease before the occurrence of the temporal break at which the
light curve starts steepening, because till then the observed evolution does
not differ from a spherically symmetric case (the beaming angle is still much
smaller than the jet opening angle), and to stabilize after the break to a
constant value (Sari et al. 1998; Sari et al. 1999).  To test this temporal
behavior, we have compared the X-ray and optical data of the GRB990510
afterglow: its simultaneous X-ray and optical light curves are shown in
Fig.~3a. The host galaxy contribution (which is very small, $V \approx 28$,
Fruchter et al. 2000; Bloom 2000) has not been subtracted from the optical
data.  We have selected seven epochs at which simultaneous measurements in
X-rays and in at least 3 optical bands are available. The simultaneity
criterion was a time separation between the optical and X-ray sampling of less
than 20 minutes;  only for May 11.978 the separation is 1.3 hr, which we have
verified to be larger than the time scale for significant optical variability
at that epoch.  We note that, since $\nu_C$ is predicted to be well above the
B optical band (PK00), the slight curvature of the optical spectrum observed
by Stanek et al. (1999) and Beuermann et al. (1999) in this band cannot be
ascribed to synchrotron cooling, and may be due instead to intrinsic or
intervening absorption.  Therefore, for consistency we excluded the B band
data from the following fit and computation. After dereddening the optical
data for the Galactic absorption using $E_{B-V} = 0.2$ mag and the extinction
curve of Cardelli et al. (1989), converting them to fluxes with the zero
points of Fukugita et al. (1995), and rescaling them, via interpolation, to
the time of the X-ray measurement, we estimated $\nu_C$ at each epoch by
intersecting the optical spectral slope (evaluated through a power-law fit of
the VRI data) and the X-ray spectral slope, as reported by KAK00 in their
Table 3.  The resulting values are reported in Fig.~3b:  $\nu_C$ is
approximately $7 \times 10^{15}$ Hz and does not exhibit significant
variability, given the large errors.

\begin{figure}
\psfig{figure=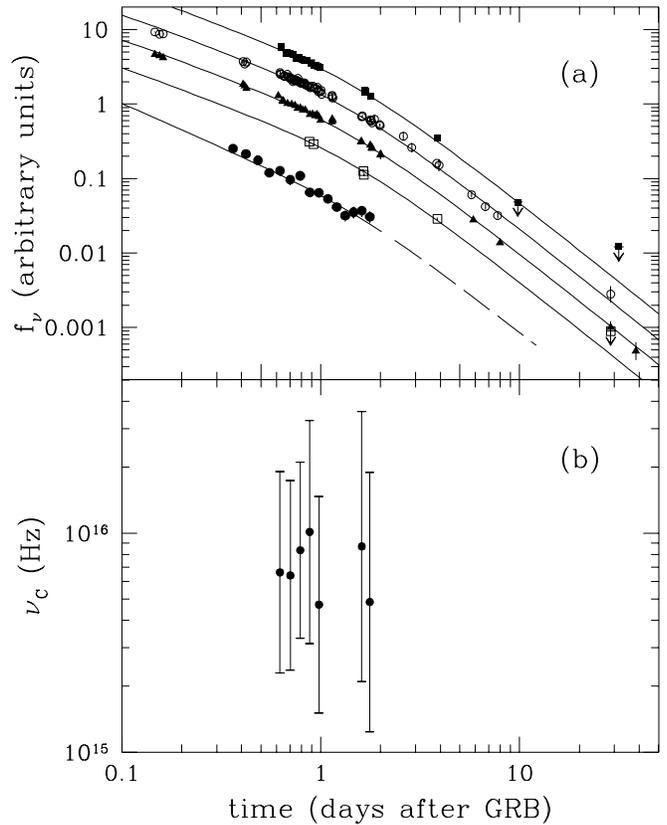,width=13cm}

\caption[]{(a) Simultaneous X-ray and optical light curves of the GRB990510
afterglow. Optical data in I (filled squares), R (empty circles), V (filled
triangles) and B (open squares) filters, and X-ray data at 5 keV (filled
circles) have been collected from Harrison et al.  (1999), Stanek et al. 
(1999), Israel et al. (1999), Beuermann et al. (1999), Fruchter et al. (1999),
KAK00. The flux normalization is in arbitrary units, except for the X-ray
data, which are reported in $\mu$Jy. Superimposed on the data are the curves
(solid)  corresponding to the model of PK00.  The model curve relative to the
X-ray data is extrapolated up to 10 days after GRB (dashed).  The model
parameters are reported in the caption to Fig.~1. (b) Synchrotron cooling
frequency, estimated from simultaneous optical and X-ray data, as a function
of time}

\end{figure}

\section{Discussion}

The prompt (1-70 s after trigger) X-ray emission of GRB990510 exhibits erratic
behavior (Fig.~1), suggestive of internal shocks occurring in an unsteady
wind, rather than an external shock propagating in the circumburst medium. 

Most of the GRB X-ray afterglows observed by BeppoSAX are well accounted for
by isotropic synchrotron radiation in an external shock and described by a
single power-law decay of index $\alpha \sim 1.1-1.6$ which connects smoothly
with the tail of the X-ray prompt emission detected by the WFC simultaneously
with the GRB (Costa et al. 1997;  Piro et al. 1998; in 't Zand et al. 1998; 
Nicastro et al. 1999; Costa 1999).  This suggests that generally the latest
X-rays detected during a GRB already represent afterglow emission (Sari \&
Piran 1999;  Frontera et al. 2000).

In GRB990510, although a single power-law $f(t)  \propto t^{-1.42}$ fits well
the X-ray afterglow data ($3 \times 10^4 - 2 \times 10^5$ seconds after
trigger, KAK00), its extrapolation backward to the time of the burst exceeds
the flux of the WFC last four points and upper limits (Fig.~1).  Therefore the
X-ray afterglow emission fall-off until at least 1000 seconds must have been
shallower than at 1 day, i.e. there was a break in the X-ray emission prior to
or during the NFI measurements. This break can be due either 1) to the passage
of a spectral break, either injection or cooling, in a spherical symmetry, or
2) to collimation of ejecta. Assuming case (1), the first WFC upper limit and
the first NFI data require that the X-ray decay prior to the spectral break
passage is shallower than $t^{-1}$.  It is unlikely that the injection break
passes through the X-ray band after 1000, as this would require $\epsilon_e$
and $\epsilon_B$ to be close to equipartition, and a large energy $E_0$ 
(then
the cooling frequency would be well below injection, and the X-ray light-curve
would decay as $t^{-1/4}$ prior to the passage of the injection break, and as
$t^{-(3p-2)/4}$ thereafter). On the other hand, the steepening of the
light-curve decay from shallower than $t^{-1}$ to $t^{-1.4}$ exceeds the
maximum index increase of 1/4 that the cooling break passage can produce.
Thus, the only remaining possibility is case (2), i.e., that the ejecta are
collimated. Furthermore, the break seen in the optical data also argues in
favor of the collimation.  Synchrotron radiation in a shock developing in a
jet which decelerates in a uniform medium accounts well for the X-ray
afterglow emission, is consistent with the last WFC measurement, and satisfies
the constraints imposed by the three WFC upper limits (Fig. 1).

The transition from the temporal decline rate in the last phase of the burst
and the early afterglow as predicted by the model is not smooth.  The final
part of the WFC profile (from 80 to 130 seconds after the GRB) is
decreasing
much more fastly ($t^{-3.7}$) than the predicted afterglow during the same
interval and soon thereafter, and therefore cannot be reconciled with it (Sari
\& Piran 1999).  An explanation may be that, even if the afterglow of
GRB990510 may start earlier than $\sim$130 s after GRB, till then the
emission
is dominated by internal shocks. This early X-ray decline can be steeper than
$2 + \beta$, where $\beta$ is the spectral index of the GRB X-ray emission
(see Fig.~1 in Kumar and Panaitescu 2000).  When the external shock starts
dominating the emission, the X-ray spectral index is expected to assume the
value $\beta = p/2 \simeq 1$, consistent with the X-ray spectral slope
measured by KAK00 for the afterglow.  However, we measure an index $\beta
\sim$1 also for the WFC spectrum during the last 50 seconds of the prompt
event (i.e., from 80 to 130 seconds after trigger), which strongly
suggests that the GRB X-ray tail already represents
afterglow emission (see e.g. the case of GRB970228, Frontera et al. 1998). 

The mismatch between the temporal decay of the observed last GRB phase and
the predicted early afterglow in GRB990510 may be removed by assuming that
the X-ray afterglow starts at the last GRB peak, observed by the WFC at 72
seconds after the instrumental trigger time. As noted by Giblin et al.
(1999) for GRB980923, there is no reason why the afterglow start time should
coincide with the time of the trigger. In fact, the internal shock GRB
emission and external shock afterglow emission are physically different and
may be independent in timing (see also GRB920723, Burenin et al. 1999).  The
new assumption of initial time allows us to describe with an external shock
model also the latest X-ray emission of the GRB, more in line with
observations of many other GRBs.  Only the first of the WFC upper limits is
clearly
inconsistent with the model (see Fig. 2).  Rather than considering this as
an evidence against a same mechanism producing the X-ray GRB tail and
afterglow, we tend to believe that the expanding blast wave crosses a region
of reduced density ambient gas, similar to what has been argued for the
X-ray afterglow of GRB981226, as reported by Frontera et al.  (2000). Those
authors conclude that this may point to a dying massive star scenario for
the GRB progenitor.  The data are insufficient to elaborate on this
hypothesis as a viable origin scenario for GRB990510, or to speculate
further on the cause of the discrepancy between estimated upper limit and
model prediction. (Note that shifting the initial time by more than 72
seconds with respect to trigger time allows recovering qualitative
consistency of the first WFC upper limit with a single power-law from the
start of the burst tail to the afterglow, but leads to exceedingly small $p$
values, and therefore does not allow a good fit of the overall
multiwavelength afterglow.)

The X-ray spectral shape during the 80-130 seconds interval after the
GRB,
$\nu^{-1}$, suggests that, in a $\nu f_\nu$ representation, the peak energy is
located within or just below the WFC band at this epoch.  Since the physical
parameters of the shock (see Fig.~1) imply that the transition between the
fast and slow cooling occurs at $\simlt$100 seconds after the GRB (taking
into
account inverse Compton losses), this peak corresponds to the crossing of the
injection ($\nu_m$) and cooling ($\nu_C$)  frequencies. The $\nu_C$
subsequently decreases, in a way approximately consistent with adiabatic or
radiative evolution, reaching the far-ultraviolet band at the collimation
break time ($\sim$1 day after GRB), as inferred from the optical and X-ray
spectral indices ($\beta_{op} \simeq 0.6$, Stanek et al. 1999; $\beta_X \simeq
1.1$, KAK00), as predicted by PK00 (see their Fig.~3), and as computed
empirically by us (Fig.~3b). Our values of $\nu_C$ after the collimation
break
time are compatible with the predicted constancy (Sari et al.  1998;  Sari et
al.  1999), although they do not provide a strong proof (the error bars for
$\nu_C$ are sufficiently large to allow the $t^{-1/2}$ behavior expected for
spherical afterglows. Moreover, if the electron cooling is inverse Compton
dominated, then $\nu_C$ may be flat in time).

\section{Conclusion}

In this paper we have focussed on the X-ray temporal behavior of
GRB990510 with the aim of confirming observationally the validity,
at these frequencies, of a decelerating and sideways expanding jet emitting
geometry. 
We have reached the following conclusions:

1) the X-ray afterglow of GRB990510 can not be described by
a single power-law,
because the extrapolation of this model back to the time of the GRB is
inconsistent with the latest X-ray data of the prompt event;

2) synchrotron radiation in a relativistic shock which propagates in a
jet
decelerating in a uniform external medium results in a continuously steepening
temporal curve, which accounts well for the multiwavelength afterglow. The
X-ray afterglow prediction at $\sim$100 seconds after trigger matches the
last
observed WFC point and is consistent with the WFC upper limits.
Thus, not only GRB990510 is the burst for which the clearest evidence of a
spreading jet has been found in the optical afterglow, but it is also the
unique case for which such evidence is present in the X-ray afterglow;

3) despite the good consistency noted in point (2) above, the temporal
slopes of prompt event latest portion and afterglow differ by $\Delta\alpha
\simeq 2.7$. 
Their similarity can be recovered if one assumes different physical
initial times for the GRB and the afterglow;

4) in this last scenario, the first WFC upper limit still suggests a
(not dramatic) deviation from a simple, monotonic behavior, which may
be ascribed to ambient medium non-uniformity. 

Future observations of multiwavelength prompt and delayed emission from GRBs
with a better temporal coverage and sampling during the first two days after
the burst than it is now affordable, will allow us to study in detail the
jet phenomenon in these events.

\acknowledgements{E. Pian is grateful for hospitality at the Space Telescope
Science Institute under the Visitor Program during completion of part of
this work. The referee, Ralph Wijers, is gratefully acknowledged for his
comments and suggestions, which led to important conclusions and contributed
to improve the paper.}


\begin{thebibliography}{}

\bibitem{} Arnaud K.A., 1996, Astronomical Data Analysis Software and
Systems V, eds. Jacoby, J. and Barnes, J., ASP Conf. Series 101, 17
 
\bibitem{} Beuermann K., Hessman F.V., Reinsch K., et al.,
1999, A\&A 352, L26

\bibitem{} Bloom J.S., 2000, GRB Circular
Notice\footnote{http://gcn.gsfc.nasa.gov/gcn/gcn3\_archive.html} N. 756

\bibitem{} Briggs M., 2000, Talk at the Conference ``GRB in the Afterglow 
Era", held in Rome (Italy), 16-20 October 2000

\bibitem{} Burenin R.A., Vikhlinin A.A., Gilfanov M.R., et al., 1999,
A\&A 344, L53

\bibitem{} Cardelli J. A., Clayton G. C., Mathis J. S., 1989, ApJ
345, 245

\bibitem{} Costa E., Frontera F., Heise J., et al., 1997, Nature 387, 783

\bibitem{} Costa E., 1999, A\&AS 138, 425

\bibitem{} Covino S., Lazzati D., Ghisellini G., et al., 1999, A\&A 348,
L1

\bibitem{} Frontera F., Costa E., Piro L., et al., 1998, ApJ 493, L67

\bibitem{} Frontera F., Amati L., Costa E., et al., 2000, ApJS 127, 59


\bibitem{} Fruchter A.S., Ferguson H., Pepper J., et al., 1999, 
GRB Circular Notice N. 386

\bibitem{} Fruchter A.S., Hook R., Pian E., 2000, GRB Circular Notice N. 
757

\bibitem{} Fukugita M., Shimasaku K., Ichikawa T., 1995, PASP 107, 945

\bibitem{} Giblin T. W., van Paradijs J., Kouveliotou C., et
al. 1999, ApJ 524, L47

\bibitem{} Harrison F., Bloom J.S., Frail D.A., et al., 1999, ApJ
523, L121

\bibitem{} Israel G.L., Marconi G., Covino S., et al., 1999, A\&A 348,
L5

\bibitem{} Jager R., Mels W.A., Brinkman A.C., et al., 1997, A\&AS 125,
557

\bibitem{} Kumar P., Panaitescu A., 2000, ApJ 541, L51

\bibitem{} Kuulkers E., Antonelli L.A., Kuiper L., et al., 2000, ApJ 538,
638 (KAK00)

\bibitem{} Nicastro L., Amati L., Antonelli L.A., et al., 1999, A\&AS 138,
437

\bibitem{} Panaitescu A., Kumar P., 2000, ApJ, submitted
(astro-ph/0010257) (PK00)

\bibitem{} Piro L., Amati L., Antonelli L.A., et al., 1998, A\&A
331, L41

\bibitem{} Rhoads J.E., 1999, ApJ 525, 737

\bibitem{} Sari R., Piran T., Narayan R., 1998, ApJ 497, L17

\bibitem{} Sari R., Piran T., Halpern J.P., 1999, ApJ 519, L17

\bibitem{} Sari R., Piran T., 1999, ApJ 520, 641

\bibitem{} Stanek K.Z., Garnavich P.M., Kaluzny J., Pych W., 
Thompson I., 1999, ApJ 522, L39

\bibitem{} Vreeswijk P.M., Fruchter A.S., Kaper L., et al., 2001,
ApJ, 546, 672

\bibitem{} Wei D.M., Lu T., 2000, A\&A, submitted (astro-ph/0012007)

\bibitem{} Wijers R.A.M.J., Vreeswijk P.M., Galama T.J., et al., 1999,
ApJ 523, L33

\bibitem{} in 't Zand J.J.M., Amati L., Antonelli L.A., et al., 1998, ApJ
505, L119

\end{thebibliography}
\end{document}